\documentclass[pdflatex,sn-mathphys-num]{sn-jnl}

\usepackage[version=4]{mhchem}
\usepackage[T1]{fontenc}
\usepackage{float} 
\usepackage{siunitx}
\usepackage{graphicx}%
\usepackage{multirow}%
\usepackage{amsmath,amssymb,amsfonts}%
\usepackage{amsthm}%
\usepackage{mathrsfs}%
\usepackage[title]{appendix}%
\usepackage{xcolor}%
\usepackage{comment}
\usepackage{textcomp}%
\usepackage{manyfoot}%
\usepackage{booktabs}%
\usepackage{algorithm}%
\usepackage{algorithmicx}%
\usepackage{algpseudocode}%
\usepackage{listings}%
\usepackage{esint}%

\theoremstyle{thmstyleone}%
\theoremstyle{thmstyletwo}%

\theoremstyle{thmstylethree}%

\begin{document}


\title{Temporal Fourier Optics Reveals Hidden Hybridized Light–Matter States}


\author[1]{\fnm{Juan-Feng} \sur{Zhu}}
\equalcont{These authors contributed equally to this work.}

\author[1]{\fnm{Wenjie} \sur{Zhou}}
\equalcont{These authors contributed equally to this work.}
\email{wenjie\_zhou@mymail.sutd.edu.sg}

\author[2,3]{\fnm{Shihao} \sur{Feng}}
\equalcont{These authors contributed equally to this work.}

\author[4,5]{\fnm{Zicheng} \sur{Song}}
\author[4,5]{\fnm{Ruicong} \sur{Zhang}}

\author[6,7]{\fnm{Cheng-Wei} \sur{Qiu}}
\email{chengwei.qiu@nus.edu.sg}

\author[2,3]{\fnm{Rengming} \sur{Liu}}
\email{liurm@henu.edu.cn}

\author[1]{\fnm{Lin} \sur{Wu}}
\email{lin\_wu@sutd.edu.sg}

\affil[1]{Science, Mathematics, and Technology, Singapore University of Technology and Design, 8 Somapah Road, 487372, Singapore}

\affil[2]{Henan Key Laboratory of High Efficiency Energy Conversion Science and Technology, Henan International Joint Laboratory of New Energy Materials and Devices, School of Physics and Electronics, Henan University, Kaifeng 475004, China}

\affil[3]{Institute of Quantum Materials and Physics, Henan Academy of Sciences, Zhengzhou 450046, China}

\affil[4]{Center for Composite Materials and Structures, Harbin Institute of Technology, Harbin 150080,China}

\affil[5]{Zhengzhou Research Institute, Harbin Institute of Technology, Zhengzhou 450018, China}

\affil[6]{Department of Electrical and Computer Engineering, National University of Singapore, 4 Engineering Drive 3, 117583, Singapore}

\affil[7]{National University of Singapore Suzhou Research Institute, Suzhou 215000, China}


\abstract{
Spectral measurements provide fundamental insights into wave systems by revealing resonances, mode hybridization, and light--matter interactions.
However, intrinsic dissipation and measurement-related spectral broadening often obscure the spectral signatures of the underlying hybridized light--matter states. Here, we establish a temporal Fourier optics framework based on a space–time Fourier correspondence, which interprets spectral broadening as the Fourier counterpart of temporal attenuation. This perspective introduces a temporal point-spread function (TPSF) that enables direct, synthesis-free reconstruction of the underlying spectral response from experimentally measured spectra by compensating for effective temporal decay before transformation back to the frequency domain. We experimentally validate the framework using deterministic single-molecule Au nanosphere dimers and open Au@Ag nanorod- and nanotriangle-based plasmonic nanocavities coupled to J-aggregate excitons. Across these distinct platforms, TPSF consistently resolves hidden upper and lower polaritonic branches, revealing hybridized light–matter states and strong coupling that remain inaccessible in conventional scattering spectra.
The reconstructed spectra agree closely with the recently developed complex-frequency formalism while providing a substantially simpler and experimentally accessible implementation. More broadly, temporal Fourier optics establishes a general framework for recovering dissipation-obscured spectral information, opening new opportunities for spectroscopy, imaging, sensing, and inverse wave measurements across photonics and wave physics.}


\maketitle

\section{Introduction}\label{sec1}

Spectroscopy provides one of the most powerful approaches for probing wave systems by revealing resonances, mode hybridization, and light--matter interactions through their spectral response \cite{weisbuch1992observation,chikkaraddy2016single,bitton2020vacuum}. In practice, however, intrinsic dissipation together with finite detection time and instrumental imperfections broaden or distort measured spectra, obscuring the underlying physical states encoded within them \cite{dorrer2000spectral,rohart2014absorption,torok2019precision,bitton2020vacuum}. Recovering these hidden states, rather than merely sharpening spectral features, remains a fundamental challenge across photonics, spectroscopy, sensing, and wave physics \cite{mazelanik2022optical,guan2023overcoming,guan2024compensating,zeng2024synthesized}.

Strong light--matter coupling provides an ideal testbed for this challenge. As a fundamental regime of cavity quantum electrodynamics, strong coupling underpins a broad range of quantum and photonic technologies, including quantum information processing \cite{imamog1999quantum,verstraete2009quantum,pan2001entanglement}, nonlinear optics \cite{chang2014quantum}, and polaritonic devices \cite{frisk2019ultrastrong}. Experimentally, it is commonly identified through Rabi splitting or avoided-crossing behaviour, where coherent coupling between a cavity mode and an emitter resonance produces upper and lower polariton branches \cite{weisbuch1992observation,yoshie2004vacuum,li2022room,park2019tip,liu2017strong,niu2022unified,baranov2018novel,zhou2025spatio}. However, the existence of hybridized light--matter states is fundamentally different from their spectroscopic observability. Within a damped coupled-mode description, the onset of strong coupling is determined by the condition
$g>(\gamma_{\rm cav}-\gamma_e)/4$,
where $g$ is the coupling strength, $\gamma_{\rm cav}$ denotes the total cavity decay rate including radiative and absorptive contributions, and $\gamma_e$ is the intrinsic emitter decay rate. In contrast, resolving two distinct spectral peaks requires a more stringent condition governed by the total linewidth together with additional measurement-induced broadening. Consequently, the absence of a clearly resolved Rabi doublet does not necessarily imply the absence of the underlying hybridized light--matter states.

Historically, experimental efforts have primarily focused on suppressing dissipation to improve spectral visibility. High-$Q$ dielectric resonators, including semiconductor microcavities \cite{weisbuch1992observation,reithmaier2004strong}, Fabry--P\'erot resonators \cite{lidzey1998strong}, micropillars \cite{wertz2010spontaneous}, photonic-crystal cavities \cite{yoshie2004vacuum,hennessy2007quantum,long2015coherent}, and whispering-gallery-mode resonators \cite{aoki2006observation,will2021coupling,vasista2020molecular}, have enabled landmark demonstrations of strong coupling. More recently, plasmonic nanocavities have achieved room-temperature strong coupling by exploiting extreme electromagnetic confinement rather than ultrahigh quality factors \cite{song2026enhanced,chikkaraddy2016single,santhosh2016vacuum,han2018rabi,liu2024deterministic}. Nevertheless, intrinsic Ohmic and radiative losses, together with finite spectral resolution, detector response, optical aberrations, and other instrumental effects, inevitably broaden measured spectra and often conceal the underlying hybridized light--matter states even when strong coupling persists \cite{dorrer2000spectral,rohart2014absorption,torok2019precision,bitton2020vacuum,khurgin2015deal,liu2025spectral,song2026enhanced}. Such limitations are not specific to strong coupling, but are common to dissipative wave systems in which intrinsic spectral features are obscured by broadening.

Here, we establish a space--time Fourier correspondence that gives rise to \emph{temporal Fourier optics}, a framework for recovering physical states hidden by spectral broadening. Inspired by the concept of the point-spread function (PSF) in spatial imaging \cite{braat2008assessment}, temporal Fourier optics reconstructs intrinsic spectral information directly from experimentally measured spectra. We demonstrate the framework by revealing hidden hybridized light--matter states in representative strong-coupling systems that remain unresolved in conventional spectroscopy. More generally, temporal Fourier optics provides a new framework for recovering dissipation-obscured information across photonics and wave physics.

\section{Temporal Fourier Optics}\label{sec2}
Temporal Fourier optics is founded on the space--time Fourier correspondence governing wave propagation. In conventional Fourier optics, a finite aperture diffracts the optical field and redistributes its spatial-frequency content. Its temporal counterpart arises naturally from the same Fourier relationship: temporal attenuation modifies the temporal response and broadens its frequency-domain spectrum. Although these phenomena are traditionally treated in different physical scenarios, they share a common mathematical origin through the Fourier transform linking physical and reciprocal domains. This correspondence motivates the temporal point-spread function (TPSF) and provides a unified framework for recovering hidden spectral information from measured spectra. In this section, we first review the spatial point-spread function (PSF), then introduce its temporal analogue and establish the underlying space--time Fourier correspondence, before developing a regularized TPSF reconstruction procedure for experimentally measured spectra.

\subsection{Space--Time Fourier Correspondence}

The mathematical correspondence between spatial diffraction and temporal spectral broadening follows from the Fourier duality between multiplication and convolution. In the spatial domain, a finite aperture acts as a spatial window, producing a broadened distribution in momentum space. In the temporal domain, finite temporal attenuation acts as a temporal window, producing an analogous broadening in the frequency domain.
This shared mathematical structure allows the PSF concept of Fourier optics to be extended naturally to the temporal domain, forming the basis of the TPSF framework.

\begin{figure*}[ht!]
\centering
\includegraphics[scale=0.75]{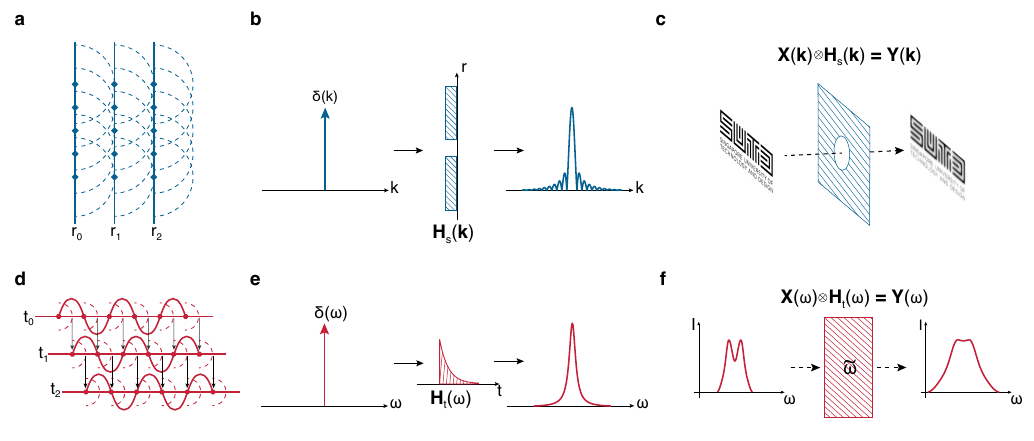}
\caption{\textbf{Space--time Fourier correspondence underlying temporal Fourier optics.}
\textbf{a,d,} Conceptual illustration of wave propagation in the spatial and temporal domains. Following the Huygens--Fresnel principle, each point on a spatial wavefront or temporal waveform contributes coherently to subsequent propagation, establishing the space--time Fourier correspondence between spatial diffraction and temporal evolution.
\textbf{b,e,} Fourier relationship between the impulse response and its corresponding transfer function. In the spatial domain, a finite aperture produces a spatial point-spread function (PSF). Its temporal analogue is the temporal point-spread function (TPSF), arising from effective temporal attenuation and describing spectral broadening induced by dissipation.
\textbf{c,} Spatial convolution with the PSF broadens each object point, resulting in image blurring and reduced spatial resolution.
\textbf{f,} Temporal convolution with the TPSF broadens intrinsic spectral features and conceals closely spaced resonances, including hybridized light--matter states. This space--time Fourier correspondence establishes the conceptual foundation of temporal Fourier optics and enables hidden spectral reconstruction in dissipative wave systems.
\label{fig1}
}
\end{figure*}

\subsection{Spatial Point--Spread Function}

We begin with the spatial description in Fourier optics, where diffraction arises from the finite extent of an optical aperture. As illustrated in Fig.~\ref{fig1}a, the Huygens--Fresnel principle describes wave propagation by treating every point on a propagating wavefront as a secondary source. The optical field at an observation point is obtained through the coherent superposition of these secondary wavelets. Wave propagation may therefore be interpreted as a continuous reconstruction of the optical wavefront, providing the physical basis for diffraction and Fourier optics.

Figure~\ref{fig1}b illustrates the same reconstruction process in Fourier space. An ideal plane wave corresponds to a delta-function distribution in transverse momentum space. After transmission through a finite aperture, the optical field becomes spatially confined and consequently broadens in transverse momentum, producing the familiar diffraction pattern. The resulting far-field response, denoted by $H_{\mathrm{s}}(\mathbf{k})$, characterizes the spatial response of the optical system.
For an aperture described by the transmission function $u_{\mathrm{s}}(\mathbf{r})$, the far-field response under the Fraunhofer condition $S\ll \lambda L$, where $S$ denotes the aperture area, $\lambda$ is the wavelength, and $L$ is the propagation distance, is given by
\begin{equation}
H_{\mathrm{s}}(\mathbf{k})
\propto
\iint_{-\infty}^{+\infty}
u_{\mathrm{s}}(\mathbf{r})
e^{i\mathbf{k}\cdot\mathbf{r}}
\,d^{2}\mathbf{r},
\label{eq1}
\end{equation}
where $\mathbf{r}$ denotes the transverse position vector within the aperture and $\mathbf{k}$ is the corresponding transverse wavevector. Throughout this work, we adopt the $\exp(-i\omega t)$ time-harmonic convention. Equation~\eqref{eq1} is the Fraunhofer diffraction integral, expressing the far-field response as the coherent superposition of the secondary wavelets transmitted through the aperture. It directly establishes the Fourier-transform relationship
\begin{equation}
H_{\mathrm{s}}(\mathbf{k})
\propto
\mathcal{F}\!\left[u_{\mathrm{s}}(\mathbf{r})\right],
\label{eq2}
\end{equation}
which forms the mathematical foundation of coherent diffraction and Fourier optics.
For an incident spatial spectrum $X(\mathbf{k})$, the finite aperture modifies the transmitted spectrum according to
$Y(\mathbf{k}) = X(\mathbf{k})
\otimes H_{\mathrm{s}}(\mathbf{k})$,
where $Y(\mathbf{k})$ denotes the transmitted spatial-frequency distribution and $\otimes$ represents convolution.

An equivalent interpretation is obtained in the image plane, where the recorded optical field is given by the convolution of the ideal image with the spatial PSF. As illustrated in Fig.~\ref{fig1}c, when an object is imaged through a finite aperture, each object point spreads into a finite diffraction pattern, causing the recorded image to appear blurred relative to the ideal object. The spatial PSF therefore describes how a finite aperture redistributes spatial-frequency components through Fourier convolution.
This spatial Fourier description also provides the reference for the temporal formulation developed in the following subsection. As will be shown, temporal attenuation broadens the frequency spectrum through an identical Fourier-transform relationship, establishing the temporal point-spread function as the natural counterpart of the spatial point-spread function.

\subsection{Temporal Point--Spread Function}
Building upon the spatial description above, an equivalent Fourier formulation can be constructed in the temporal domain. In this temporal analogue, effective attenuation plays a role analogous to that of a finite spatial aperture, producing spectral broadening through Fourier transformation. This correspondence naturally gives rise to the TPSF, which characterizes the influence of temporal attenuation on the measured spectral response.

The temporal counterpart of the Huygens--Fresnel construction is illustrated schematically in Fig.~\ref{fig1}d. Similar to the continuous reconstruction of a propagating wavefront in the spatial domain, the waveform at time $t_{0}$ contributes to the reconstructed waveform at a later time $t_{1}$, which subsequently contributes to the field at $t_{2}$. Temporal evolution may therefore be viewed as a continuous reconstruction process, providing the temporal analogue of wave propagation.
The corresponding Fourier description is illustrated in Fig.~\ref{fig1}e. An ideal monochromatic response is represented by a delta-function distribution in frequency domain. When the temporal response undergoes homogeneous attenuation, it is multiplied by a temporal attenuation function, which broadens the corresponding spectral response after Fourier transformation. The resulting frequency-domain response defines the TPSF, denoted by $H_{\mathrm{t}}(\omega)$.
For a temporal attenuation function $u_{\mathrm{t}}(t)$, the TPSF is given by
\begin{equation}
H_{\mathrm{t}}(\omega)
\propto
\int_{-\infty}^{+\infty}
u_{\mathrm{t}}(t)
e^{-i\omega t}
\,dt,
\label{eq3}
\end{equation}
which immediately establishes the Fourier-transform relationship
\begin{equation}
H_{\mathrm{t}}(\omega)
\propto
\mathcal{F}\!\left[u_{\mathrm{t}}(t)\right].
\label{eq4}
\end{equation}
Equation~\eqref{eq4} is the direct temporal counterpart of Eq.~\eqref{eq2}. Together, they establish the common Fourier description that links spatial diffraction and temporal spectral broadening within the TPSF framework.
For homogeneous causal dissipation, the temporal attenuation function is approximated as
\begin{equation}
u_{\mathrm{t}}(t)
=
e^{-\gamma(t-t_{\mathrm{start}})}
\Theta(t-t_{\mathrm{start}}),
\label{eq:causal_decay}
\end{equation}
where $\gamma$ denotes the effective temporal decay rate and $\Theta(t)$ is the Heaviside step function.
For a finite observation interval
$t_{\mathrm{start}}\le t\le t_{\mathrm{end}}$,
Eq.~\eqref{eq:causal_decay} can equivalently be written as
\begin{equation}
u_{\mathrm{t}}(t)
=
\begin{cases}
e^{-\gamma(t-t_{\mathrm{start}})},
&
t_{\mathrm{start}}\leq t\leq t_{\mathrm{end}},\\
0,
&
\text{otherwise}.
\end{cases}
\label{eq:finite_temporal_filter}
\end{equation}
The Fourier transform of the exponentially decaying temporal response gives a complex Lorentzian spectral amplitude, whose squared magnitude exhibits the familiar Lorentzian lineshape (Fig. \ref{fig1}e). Consequently, temporal attenuation redistributes spectral energy through Fourier broadening in exactly the same manner that a finite aperture redistributes transverse momentum through spatial diffraction.

For an intrinsic temporal response $x(t)$, temporal attenuation produces the measured response
$ y(t) = x(t)u_{\mathrm{t}}(t)$.
According to the convolution theorem, the corresponding measured spectrum satisfies
$ Y(\omega) = X(\omega) \otimes
H_{\mathrm{t}}(\omega)$,
where $X(\omega)$ and $Y(\omega)$ denote the intrinsic and measured spectra, respectively.
As illustrated in Fig.~\ref{fig1}f, convolution with the TPSF progressively broadens neighbouring resonances, causing their spectral overlap and eventually obscuring their individual identities. Consequently, spectral features that are clearly distinguishable in the intrinsic response may become indistinguishable in the experimentally measured spectrum. Importantly, the disappearance of resolvable spectral peaks does not imply the disappearance of the underlying physical states, but instead reflects the Fourier broadening introduced by temporal attenuation. This observation forms the physical basis for reconstructing the intrinsic spectrum through inversion of the TPSF, as developed in the following subsection.

\subsection{Unified Space–Time Fourier Formalism}

The spatial and temporal descriptions developed in the preceding sections are unified by the scalar Helmholtz  equation, which provides the mathematical foundation for the space--time Fourier correspondence underlying the TPSF framework. By separating the spatial and temporal variables, identical Fourier structures naturally emerge for spatial diffraction and temporal spectral broadening.
The scalar wave equation is given by
\begin{equation}
\nabla^{2}u(\mathbf{r},t)
-
\frac{1}{c^{2}}
\frac{\partial^{2}u(\mathbf{r},t)}
{\partial t^{2}}
=
0,
\label{eq:wave_equation}
\end{equation}
where $u(\mathbf{r},t)$ denotes the scalar wave field and $c$ is the speed of light in the medium.
Seeking separable solutions of the form
\begin{equation}
u(\mathbf{r},t)
=
U(\mathbf{r})T(t),
\label{eq:separation}
\end{equation}
naturally yields the independent spatial and temporal equations,
\begin{equation}
\left(
\nabla^{2}
+
k^{2}
\right)
U(\mathbf{r})
=
0,
\label{eq:spatial_helmholtz}
\end{equation}
and
\begin{equation}
\left(
\frac{\partial^{2}}
{\partial t^{2}}
+
\omega^{2}
\right)
T(t)
=
0,
\label{eq:temporal_oscillator}
\end{equation}
where $\omega=ck$ in a homogeneous nondispersive medium.

\begin{table*}[ht]
\caption{\textbf{Space--time Fourier correspondence underlying the temporal point-spread-function framework.}}
\label{tab1}
\centering
\begin{tabular}{ccc}
\hline\hline

&
\textbf{Spatial domain}
&
\textbf{Temporal domain}
\\
\hline

Physical process
&
Diffraction
&
Spectral broadening
\\[8pt]

Physical quantity
&
Aperture function
$u_{\mathrm{s}}(\mathbf r)$
&
Temporal attenuation
$u_{\mathrm{t}}(t)$
\\[12pt]

Governing equation
&
$\left(\nabla^{2}+k^{2}\right)U(\mathbf r)=0$
&
$\left(\dfrac{\partial^{2}}{\partial t^{2}}+\omega^{2}\right)T(t)=0$
\\[12pt]
Point-spread function
&
$
H_{\mathrm s}(\mathbf k)
=
\mathcal F
\!\left[
u_{\mathrm s}(\mathbf r)
\right]
$
&
$
H_{\mathrm t}(\omega)
=
\mathcal F
\!\left[
u_{\mathrm t}(t)
\right]
$
\\[12pt]

Fourier relationship
&
$
Y(\mathbf k)
=
X(\mathbf k)
\otimes
H_{\mathrm s}(\mathbf k)
$
&
$
Y(\omega)
=
X(\omega)
\otimes
H_{\mathrm t}(\omega)
$
\\[12pt]
Observable effect
&
Image blur
&
Hidden spectrum
\\[8pt]

Inverse operation
&
Image deconvolution
&
TPSF reconstruction
\\
\hline\hline
\end{tabular}
\end{table*}

The spatial Helmholtz equation governs diffraction in physical space, whereas the temporal harmonic-oscillator equation describes the corresponding temporal evolution. Although the governing variables differ, both equations possess identical Fourier representations relating the physical domain to its reciprocal domain. Consequently, the Fourier transform between the aperture function and the spatial PSF has a direct temporal counterpart relating the temporal attenuation function to the TPSF. For clarity, the essential physical and mathematical correspondences between the spatial and temporal descriptions are summarized in Table~\ref{tab1}. Together, they establish a unified Fourier framework that underpins the TPSF formulation for hidden spectral reconstruction.

Viewed from this perspective, hidden spectral reconstruction is the temporal analogue of image deconvolution: just as spatial deconvolution compensates the broadening introduced by a spatial PSF, TPSF reconstruction compensates the spectral broadening introduced by temporal dissipation. This analogy naturally leads to the inverse reconstruction procedure described in the following subsection.

\subsection{Regularized Temporal Response Reconstruction}

Within the TPSF framework established above, the measured spectrum is first transformed into its temporal representation, allowing the effective attenuation to be compensated in the time domain rather than through direct frequency-domain deconvolution.
For an intrinsic temporal response $x(t)$, temporal attenuation produces the measured response
$y(t)=x(t)u_{\mathrm{t}}(t)$.
In the absence of noise, the intrinsic response can formally be recovered through
$x(t)=y(t)/u_{\mathrm{t}}(t)$.
For the exponential attenuation model introduced in Eq.~\eqref{eq:causal_decay},
$ u_{\mathrm{t}}(t)=e^{-\gamma t}$,
this inverse operation is equivalent to compensating the temporal attenuation with an exponential gain factor,
$\tilde{x}(t)=y(t)e^{\gamma t}$,
where $\tilde{x}(t)$ denotes the reconstructed temporal response. The corresponding reconstructed spectrum is subsequently obtained through
$\tilde{X}(\omega)
= \mathcal{F} \!\left[\tilde{x}(t)
\right]$,
where $\tilde{X}(\omega)$ represents the reconstructed intrinsic spectrum.
Accordingly, the TPSF reconstruction procedure can be summarized as
\begin{equation}
Y(\omega)
\;
\xrightarrow{\mathcal{F}^{-1}}
\;
y(t)
\;
\xrightarrow{\times\,e^{\gamma t}}
\;
\tilde{x}(t)
\;
\xrightarrow{\mathcal{F}}
\;
\tilde{X}(\omega),
\label{eq:workflow}
\end{equation}

In practice, direct temporal compensation amplifies noise at late times because the exponential gain increases as the signal decays. 
To avoid this instability, we subsequently introduce a Wiener-regularized compensation function that suppresses noise amplification while retaining the recoverable temporal information, as described in Section~\ref{sec3}. The resulting reconstruction suppresses dissipation-induced spectral broadening while preserving the intrinsic resonance frequencies, thereby recovering spectral features that remain hidden in the experimentally measured spectrum.

\begin{figure*}[ht!]
\centering
\includegraphics[scale=0.75]{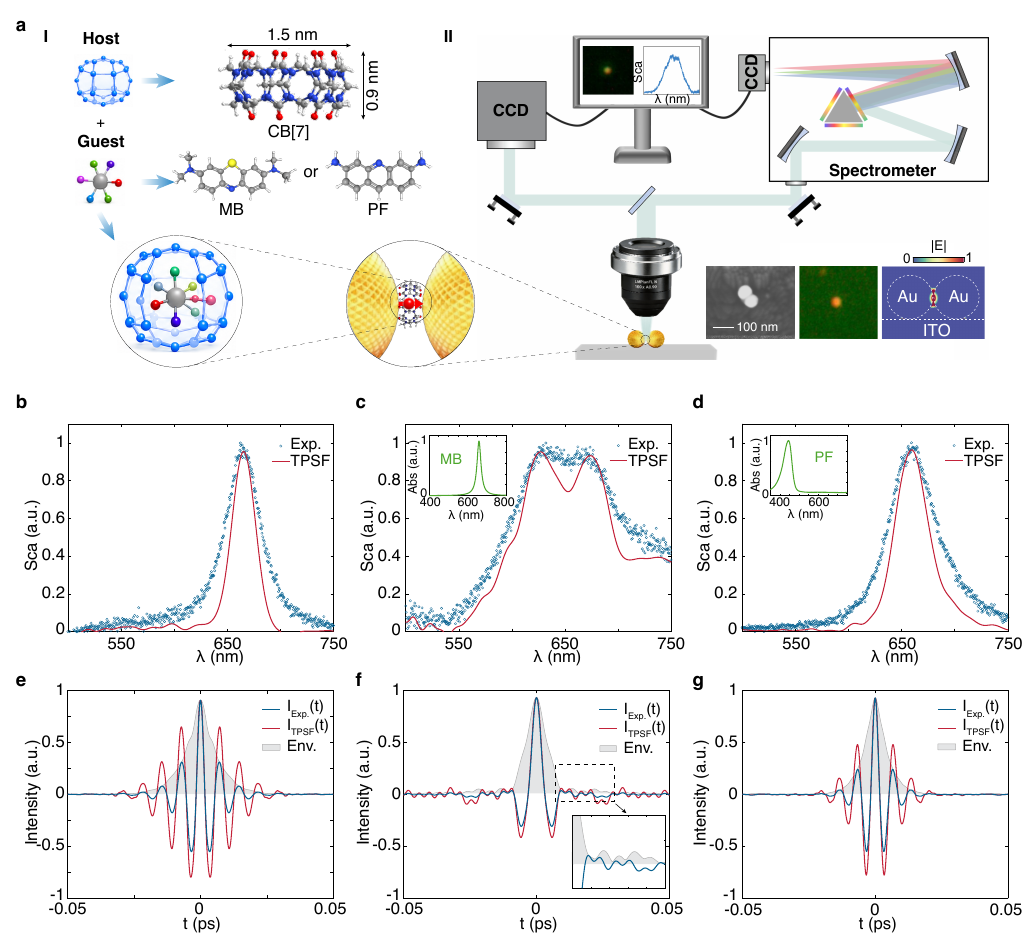}
\caption{\textbf{Experimental validation of TPSF in deterministic single-molecule plasmon--exciton nanocavities.}
\textbf{a,} Schematic of the CB$[7]$-mediated Au nanosphere dimer supporting deterministic single-molecule plasmon--exciton coupling and the corresponding dark-field scattering spectroscopy setup. Insets show a representative SEM image, measured dark-field scattering spot, and simulated electric-field distribution.
\textbf{b--d,} Experimental scattering spectra before (blue) and after TPSF reconstruction (red). Bare Au nanosphere dimers (\textbf{b}) exhibit linewidth narrowing without spectral splitting. For a single methylene blue (MB) molecule (\textbf{c}), TPSF resolves the hidden upper and lower polariton branches that remain obscured in the measured spectrum. In contrast, a spectrally detuned proflavine (PF) molecule (\textbf{d}) retains a single-resonance response after reconstruction.
\textbf{e--g,} Corresponding temporal responses. Gray shading indicates the temporal envelope. The bare Au nanosphere dimer (\textbf{e}) exhibits monotonic temporal decay, while the MB-coupled nanocavity (\textbf{f}) shows pronounced temporal beating arising from coherent plasmon--exciton hybridization. The PF control (\textbf{g}) exhibits neither temporal beating nor spectral splitting, confirming that TPSF faithfully reconstructs hidden spectral information without generating artificial strong-coupling signatures.
}
\label{fig2}
\end{figure*}

\section{Revealing Hidden Hybridized Light–Matter States at the Single-Molecule Limit}
\label{sec3}

Having established the theoretical foundation of temporal Fourier optics, we now experimentally validate the TPSF framework using one of the most demanding nanophotonic platforms: room-temperature single-molecule plasmon--exciton coupling. For an ensemble of $N$ emitters participating in the interaction, the collective coupling strength scales as $\sqrt{N}$.
At the single-molecule limit, this collective enhancement is absent, while substantial Ohmic and radiative losses in plasmonic nanocavities broaden the spectral resonances.
Consequently, the upper and lower polaritonic branches may remain indistinguishable in conventional scattering spectra even when the underlying plasmon--exciton hybridization persists. This system therefore provides a stringent test of whether TPSF can recover hidden spectral information directly from experimentally broadened spectra.
For experimental validation, we employ a deterministic single-molecule plasmonic nanocavity.
We first examine an empty-cavity reference to verify that TPSF compensates spectral broadening without introducing artificial resonances.
We then demonstrate the recovery of hidden hybridized light--matter states in the single-molecule strong-coupling regime and perform detuned control experiments to confirm that the reconstructed spectra recover information already encoded in the measured response rather than generating spurious spectral features.


\subsection{Experimental Platform}
The experimental platform is illustrated in Fig.~\ref{fig2}a. The nanocavity is constructed using the cucurbit[7]uril (CB$[7]$)-mediated host--guest assembly reported in our previous work \cite{liu2024deterministic}.
CB$[7]$ serves a dual function in this platform: its molecular cavity can encapsulate a single guest molecule, while its carbonyl portals bind to two Au nanospheres and mediate their assembly into an Au nanosphere dimer (AuND) with a reproducible sub-nanometre gap of approximately 0.9~nm. Control absorption measurements further show that CB$[7]$ contributes negligibly to the optical response in the spectral range of interest (Fig.~\textcolor{blue}{S1}).
This configuration simultaneously defines the plasmonic gap and positions the encapsulated guest molecule within the localized gap-plasmon hotspot.
The host--guest geometry further constrains the molecular transition dipole along the opening direction of the CB$[7]$ cavity, promoting both spatial and orientational overlap with the localized gap-plasmon field.
Individual AuNDs were subsequently characterized by dark-field scattering spectroscopy, as illustrated in panel II of Fig.~\ref{fig2}a. Details of the sample preparation and optical characterization are provided in the \textcolor{blue}{Methods}.

\subsection{Empty-Cavity Reference}

We first establish an empty-cavity reference to verify that TPSF compensates spectral broadening without introducing artificial resonances. AuNDs with sphere diameters of approximately 40~nm were assembled using the CB$[7]$-mediated strategy described above, but without incorporating molecular excitons. A representative scanning electron micrograph and the corresponding dark-field scattering spot are shown in the insets of Fig.~\ref{fig2}a, while the measured scattering spectrum is presented in Fig.~\ref{fig2}b. The spectrum exhibits a broad Lorentzian resonance centered near 660~nm, corresponding to the fundamental localized surface plasmon mode. 
The simulated near-field distribution at this wavelength, shown in the right inset of Fig.~\ref{fig2}a, reveals strong electromagnetic confinement within the sub-nanometer gap, with a maximum electric-field enhancement of $|E/E_0|\approx 330$, providing favorable conditions for subsequent single-molecule coupling.

Following the TPSF procedure developed in Section~\ref{sec2}, the measured scattering spectrum is transformed into the temporal domain, as shown by the blue curve in Fig.~\ref{fig2}e. The temporal representation exhibits a nearly single-frequency oscillation, while its envelope, extracted using the Hilbert transform and indicated by the light-gray shaded region, decays approximately exponentially, reflecting the finite effective decay time associated with the broadened plasmonic resonance. 
To compensate for the effective temporal attenuation while suppressing noise amplification, we introduce a Wiener-regularized inverse filter.
For the two-sided temporal representation obtained from the Fourier transform of the measured spectrum, the effective loss kernel is taken as $L(t)=e^{-\gamma|t|}$. The corresponding inverse filter is defined as
\begin{equation}
W(t) = \frac{L^{*}(t)}
{|L(t)|^{2}+\eta}
= \frac{e^{-\gamma|t|}}
{e^{-2\gamma|t|}+\eta},
\label{eq:wiener_filter}
\end{equation}
where $\gamma$ is the effective decay rate and $\eta=1/{\rm SNR}$ is the regularization parameter. Unless otherwise specified, $\mathrm{SNR}=100$ is adopted throughout this work.

The reconstructed temporal response is shown by the red curve in Fig.~\ref{fig2}e. Compared with the measured signal, the reconstructed response exhibits a substantially slower temporal decay while preserving the same oscillation frequency. After transformation back into the frequency domain, the resonance linewidth decreases from 137~meV to 72.9~meV, whereas the resonance position remains essentially unchanged, as shown in Fig.~\ref{fig2}b.
These observations demonstrate that TPSF compensates dissipation-induced spectral broadening without introducing additional resonant modes or shifting the resonance frequency. The empty-cavity system therefore establishes a reference for interpreting the subsequent reconstruction of the single-molecule strong-coupling system.

\subsection{Recovering Hidden Single-Molecule Rabi Splitting}

Having established the empty-cavity reference, we next examine the more demanding single-molecule strong-coupling regime. A single methylene blue (MB) molecule is encapsulated within the CB$[7]$ host to form a CB$[7]$@MB complex. The absorption spectrum of the complex, shown in the inset of Fig.~\ref{fig2}c, exhibits a resonance near 660~nm, closely matching the localized plasmon resonance of the AuND and thereby providing favorable conditions for plasmon--exciton hybridization.
Despite this spectral matching, the experimentally measured scattering spectrum exhibits only a broadened and flattened resonance profile, as shown by the blue symbols in Fig.~\ref{fig2}c. The upper and lower polariton branches cannot be clearly distinguished, preventing direct identification of Rabi splitting from the measured spectrum alone. This system therefore represents an ideal benchmark for evaluating whether TPSF can recover spectral information that remains hidden in the broadened response.

The temporal representation obtained from the measured spectrum is shown by the blue curve in Fig.~\ref{fig2}f. Compared with the empty-cavity reference (Fig.~\ref{fig2}e), the temporal signal exhibits a weak but discernible beating pattern, while the envelope extracted by the Hilbert transform deviates from the single-exponential decay observed for the uncoupled cavity. These features indicate the presence of multiple frequency components that are only weakly resolved in the measured  spectrum.
Applying the TPSF reconstruction procedure reveals a much more pronounced temporal beating, as shown by the red curve in Fig.~\ref{fig2}f. The slower temporal decay reveals the coexistence of two oscillation frequencies that become clearly distinguishable after temporal compensation. Following Fourier transformation back to the 
frequency domain, two well-resolved polaritonic peaks emerge at 627.4~nm and 673.9~nm, respectively, as shown by the red curve in Fig.~\ref{fig2}c.
The recovered energy splitting is $
\Omega_R=136~\mathrm{meV}$,
corresponding to a coupling strength of
$g=\Omega_R/2=68~\mathrm{meV}$
under the near-resonant condition. This value agrees well with the theoretical prediction for the deterministic single-molecule plasmon--exciton system \cite{liu2024deterministic}, providing independent validation of the reconstructed spectrum.

Importantly, TPSF enhances oscillatory components in the temporal representation of the measured spectrum without introducing new characteristic frequencies.
The resulting temporal beating and reconstructed Rabi splitting provide mutually consistent signatures of the same underlying plasmon--exciton hybridization.
This internal consistency supports the interpretation that the hybridized spectral response persists even when its polaritonic branches remain unresolved in the directly measured scattering spectrum.

\subsection{Further Validation of the TPSF Framework}

We next assess the TPSF framework under two complementary conditions beyond the near-resonant single-molecule system: a molecular control with large plasmon--exciton detuning and coupled systems at finite detuning.
These tests examine the selectivity of the reconstruction and its ability to recover coupling-dependent spectral responses under different resonance conditions.


As a negative molecular control, MB was replaced by proflavine (PF), which is chemically and structurally similar to MB and can be encapsulated within CB$[7]$ through the same host--guest mechanism. Its measured absorption resonance is centered near 445~nm, as shown in the inset of Fig.~\ref{fig2}d. Owing to the large spectral detuning between the molecular transition and the localized plasmon resonance, appreciable plasmon--exciton hybridization is not expected. Consistent with this expectation, the measured scattering spectrum exhibits a single Lorentzian resonance (blue symbols in Fig.~\ref{fig2}d), while its temporal representation (blue curve in Fig.~\ref{fig2}g) closely resembles that of the empty cavity (Fig.~\ref{fig2}e), with no pronounced beating. Applying the TPSF reconstruction narrows the resonance linewidth from 149~meV to 103~meV, demonstrating effective compensation of spectral broadening. However, neither temporal beating nor spectral splitting emerges after reconstruction. In contrast to the near-resonant MB system, the PF control remains a single-resonance system throughout the reconstruction process. These observations confirm that TPSF does not generate artificial polaritonic branches when no underlying mode splitting is encoded in the measured response.

The applicability of TPSF away from exact resonance is further examined by tuning the plasmon resonance through variation of the AuNDs (Fig.~\textcolor{blue}{S2}). In both red- and blue-detuned configurations, TPSF consistently reconstructs the corresponding upper and lower polariton branches while preserving the expected detuning dependence of the mode splitting. These additional measurements demonstrate that the framework remains effective over a range of spectral detunings rather than being restricted to the near-resonant case.

\section{Generalization to Open Plasmonic Nanocavities}
\label{sec4}


We next examine the applicability of temporal Fourier optics to open plasmonic nanocavities containing distributed excitonic ensembles. Specifically, we study Au@Ag nanorod-- and nanotriangle--J-aggregate systems, where distributed excitons couple to open plasmonic resonances. 
Compared with deterministic AuND nanogap cavities, these plasmonic cavities exhibit larger mode volumes and stronger radiative leakage, resulting in weaker field confinement and broader resonances, respectively.
These characteristics make the polaritonic features more difficult to resolve and therefore provide a stringent test of the robustness and generality of the TPSF framework.

\begin{figure*}[ht!]
\centering
\includegraphics[width=1\linewidth]{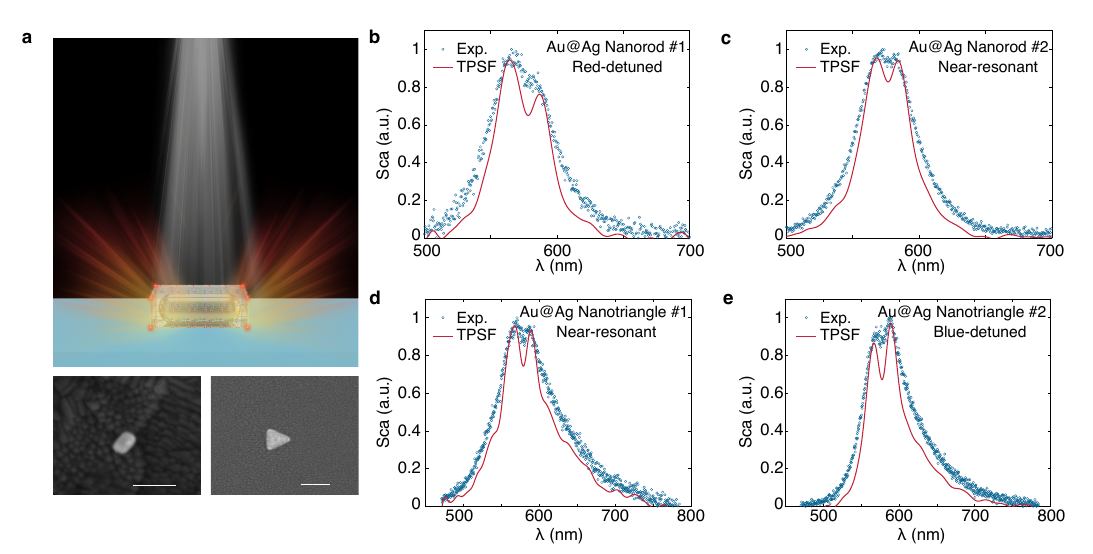}
\caption{\textbf{Generalization of the TPSF framework to realistic open plasmonic nanocavities.}
\textbf{a,} Schematic of Au@Ag nanorod- and nanotriangle-based plasmonic nanocavities coated with a thin J-aggregate layer. Insets show representative SEM images (scale bars: 100~nm).
\textbf{b--e,} Experimental scattering spectra (blue symbols) before reconstruction and the corresponding TPSF-reconstructed spectra (red curves) for representative Au@Ag nanorod (\textbf{b,c}) and nanotriangle (\textbf{d,e}) nanocavities. In all cases, TPSF recovers the hidden upper and lower polariton branches that remain obscured in the measured spectra, demonstrating that the framework is robust across different plasmonic architectures, distributed excitonic systems, and coupling conditions.
} \label{fig3}
\end{figure*}

As illustrated in Fig.~\ref{fig3}a, the Au@Ag nanorod and nanotriangle nanocavities consist of an Au core surrounded by an Ag shell. The core--shell nanostructures were deposited on an ITO substrate and subsequently coated with an approximately 0.9-nm-thick J-aggregate layer through solution-based self-assembly, as schematically illustrated in Fig.~\textcolor{blue}{S3} \cite{liu2017strong,li2023highly}.
Unlike the AuND, in which the electromagnetic field is predominantly confined within the sub-nanometre gap, the Au@Ag nanostructures support localized plasmonic modes concentrated near the sharp corners of the Ag shell.
Compared with the plasmonic mode supported by the Au core alone, the Ag shell enhances field localization and reduces the effective mode volume, thereby promoting coupling to the surrounding J-aggregate excitons \cite{liu2017strong}.
In addition, these open plasmonic structures interact with a spatially distributed excitonic layer rather than a deterministically positioned single emitter, providing a complementary platform for evaluating the applicability of TPSF to ensemble-coupled plasmonic systems. The corresponding field distribution and additional structural details are provided in Figs. ~\textcolor{blue}{S4-S5}, while fabrication and optical-characterization procedures are described in the \textcolor{blue}{Methods}.

Representative TPSF reconstructions for the Au@Ag nanorod platform are shown in Figs.~\ref{fig3}b,c. For Sample$~\#1$ (Fig.~\ref{fig3}b), the measured scattering spectrum exhibits only a broadened resonance with weak indications of mode splitting, preventing unambiguous identification of the underlying polaritonic branches using conventional spectral analysis. Following TPSF reconstruction, however, two well-resolved polariton modes emerge under the red-detuned condition, yielding a mode splitting of $87~\mathrm{meV}$, in good agreement with previously reported coupling strengths for Au@Ag nanorod--J-aggregate systems \cite{liu2017strong}. A second nanorod sample with a different resonance condition was examined to test the robustness of the reconstruction.
As shown in Fig.~\ref{fig3}c, the measured spectrum displays a broad, weakly structured profile without clearly distinguishable upper and lower polariton branches. Nevertheless, TPSF successfully reconstructs the corresponding polaritonic resonances centered at 568.1~nm and 584.5~nm. These results demonstrate that the reconstruction remains effective across different nanorod geometries and detuning conditions.

The generality of the framework was further examined using Au@Ag nanotriangle nanocavities, whose modal characteristics differ substantially from those of the nanorod platform. Representative experimental results are presented in Figs.~\ref{fig3}d,e. Although no clearly resolved mode splitting is visible in the measured spectra, TPSF resolves the corresponding upper and lower polaritonic branches for both nanotriangle samples. The recovered splittings are 79.4~meV and 83.6~meV, respectively. 
These values lie near the crossover between intermediate and strong coupling \cite{liu2021relativity,li2023highly}, where dissipation makes the polaritonic branches particularly difficult to distinguish spectroscopically.
These results further demonstrate that the effectiveness of TPSF does not depend on a particular cavity geometry or coupling regime.

\section{Discussion}\label{sec5}

Having established TPSF across distinct plasmon--exciton platforms, we now examine its robustness, physical interpretation, and broader implications for dissipative wave systems.

\subsection{Robustness of TPSF Reconstruction}

\begin{figure*}[h]
\centering
\includegraphics[scale=1]{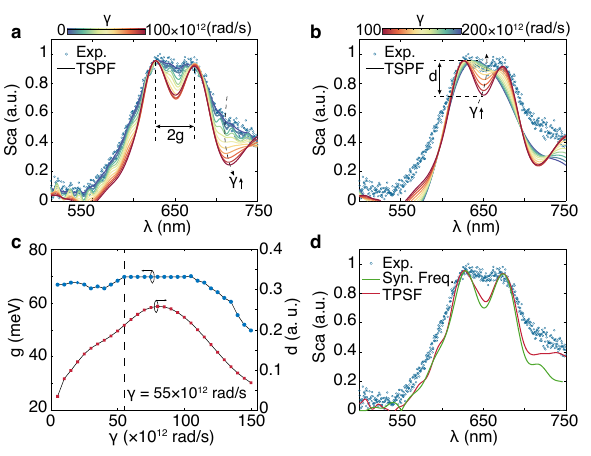}
\caption{\textbf{Robustness and physical validation of the TPSF framework.}
\textbf{a,b,} Evolution of the reconstructed scattering spectra as the compensation factor $\gamma$ increases. Progressive compensation of temporal attenuation suppresses dissipation-induced spectral broadening and reveals the hidden upper and lower polariton branches over a broad range of $\gamma$.
\textbf{c,} Recovered coupling strength $g$ (blue) and splitting depth $d$ (red) as functions of $\gamma$. The broad plateau in $g$ demonstrates that the recovered Rabi splitting is insensitive to the precise choice of $\gamma$, while overcompensation at large $\gamma$ gradually introduces reconstruction artifacts.
\textbf{d,} Comparison between the experimentally measured spectrum (blue symbols), TPSF reconstruction (red), and synthetic-frequency reconstruction (green). The excellent agreement between the two reconstruction approaches confirms that TPSF faithfully recovers the hidden spectral information encoded in the measured response.
}
\label{fig4}
\end{figure*}

A practical reconstruction framework should not depend on precise parameter tuning, but should recover the underlying spectral response over a broad range of physically reasonable conditions. We therefore examine the robustness of TPSF with respect to the effective compensation parameter $\gamma$, which phenomenologically captures the cumulative temporal attenuation and spectral broadening associated with intrinsic dissipation, structural imperfections, and measurement-related effects that are difficult to quantify independently. Using the resonant AuND--MB system as a representative example, Figs.~\ref{fig4}a,b present the reconstructed spectra obtained by varying $\gamma$ from $0$ to $100\times10^{12}\,\mathrm{rad/s}$. As $\gamma$ increases, temporal attenuation is progressively compensated, resulting in continuous narrowing of the reconstructed resonance linewidth and increasingly clear separation of the upper and lower polariton branches. The Rabi splitting becomes stably resolved once $\gamma$ exceeds approximately $55\times10^{12}\,\mathrm{rad/s}$. Meanwhile, the splitting depth $d$ continues to increase and reaches its maximum value near $\gamma=80\times10^{12}\,\mathrm{rad/s}$, corresponding to the highest visibility of the reconstructed polaritonic branches.

Importantly, although the spectral contrast improves as temporal attenuation is progressively compensated, the recovered polaritonic peak positions remain essentially unchanged over a broad range of compensation factors. As shown in Fig.~\ref{fig4}c, the extracted coupling strength $g$ rapidly converges and subsequently remains nearly constant, demonstrating that TPSF primarily compensates dissipation-induced spectral broadening while preserving the intrinsic resonance frequencies. The recovered Rabi splitting therefore reflects spectral information already encoded in the measured response rather than a consequence of parameter tuning. The broad plateau observed in Fig.~\ref{fig4}c further indicates that the reconstruction is insensitive to moderate variations of $\gamma$, providing a robust operating region for practical implementation.

When $\gamma$ becomes excessively large, however, the reconstruction begins to exhibit nonphysical distortions. As illustrated in Fig.~\ref{fig4}a, narrow spectral features gradually emerge near 720~nm when the temporal compensation exceeds the physically meaningful range. Their origin can be understood from the temporal-domain reconstruction process: excessive compensation amplifies weak late-time spectral components together with numerical fluctuations, producing artificial structures in the reconstructed spectrum. As $\gamma$ is further increased from $100\times10^{12}$ to $200\times10^{12}\,\mathrm{rad/s}$, the reconstructed spectrum gradually evolves back toward a broadened Lorentzian-like profile (Fig.~\ref{fig4}b), indicating that overcompensation ultimately degrades the reconstruction quality. Considering both the stability of the recovered coupling strength and the absence of reconstruction artifacts, $\gamma=55\times10^{12}\,\mathrm{rad/s}$ is adopted for the AuND--MB system. This value corresponds to the smallest compensation factor that consistently resolves the hidden polaritonic branches while avoiding overcompensation-induced distortions.

As an independent validation, the same $\gamma$-dependent TPSF reconstruction was applied to the PF-coupled AuND (Fig.~\textcolor{blue}{S6}a), for which strong plasmon--exciton coupling is not expected. In contrast to the resonant MB system, no Rabi splitting emerges throughout the entire range of compensation factors investigated. Instead, the reconstructed spectra consistently retain a single Lorentzian-like resonance whose linewidth decreases progressively with increasing $\gamma$. 
Together with the stability plateau observed for the MB-coupled system, this result demonstrates that the reconstructed Rabi splitting is robust against moderate variations in $\gamma$ and is not generated by parameter tuning.


\subsection{Physical Interpretation and Benchmarking}

The successful reconstruction of hidden polaritonic branches across multiple nanophotonic platforms suggests that TPSF is not merely a numerical deconvolution procedure, but reflects a deeper physical description of dissipative wave systems. 
This interpretation becomes particularly clear when the reconstruction is viewed from the perspective of complex-frequency excitation \cite{kim2025complex}. As established in Section~\ref{sec2}, compensating temporal attenuation by multiplying the measured temporal response by $e^{\gamma t}$ is mathematically equivalent to translating the system response into the complex-frequency plane through the transformation $\omega\rightarrow\omega-i\gamma$ \cite{kim2025complex}. TPSF therefore provides an alternative route for accessing complex-frequency responses directly from experimentally measured real-frequency spectra.
 
Conventionally, a direct implementation of complex-frequency excitation requires a temporally modulated waveform with an exponentially growing or decaying envelope \cite{gu2022transient,hinney2024efficient,xu2025space,xue2026lasing}. Although physically direct, such excitation generally requires precise control over the temporal amplitude and phase and can therefore be challenging to implement experimentally. More recently, synthetic-frequency methods have provided an elegant alternative by synthesizing broadband real-frequency responses to emulate complex-frequency excitation without explicit temporal waveform engineering \cite{guan2023overcoming,guan2024compensating,wu2025physics,zeng2024synthesized,guan2026high,zhang2026loss}. This concept has been successfully applied to a broad range of platforms, including super-resolution imaging \cite{guan2023overcoming}, hyperbolic polariton propagation \cite{guan2024compensating}, sensing \cite{zeng2024synthesized,wu2025physics}, and electron energy-loss spectroscopy \cite{chen2026synthetic}.

From this perspective, TPSF and synthetic-frequency reconstruction represent two distinct numerical routes for accessing the same underlying complex-frequency response and are functionally equivalent in the systems examined here.
To verify this equivalence, the synthetic-frequency approach was applied to the near-resonant AuND--MB system. As shown by the green curve in Fig.~\ref{fig4}d, synthetic-frequency reconstruction recovers well-resolved upper and lower polariton branches that agree remarkably well with the TPSF reconstruction (red curve). The two approaches yield nearly identical polaritonic peak positions and coupling strengths, demonstrating that they recover the same hidden spectral information encoded in the experimentally measured response within numerical accuracy.
Further validation is provided by the PF-coupled AuND. When synthetic-frequency reconstruction is applied to this detuned system, the reconstructed spectrum remains a narrowed Lorentzian resonance without observable Rabi splitting (Fig.~\textcolor{blue}{S6}b), fully consistent with the TPSF result. This agreement across the coupled and control systems further supports the physical consistency of the TPSF reconstruction.

Although both approaches access complex-frequency information and yield equivalent reconstruction outcomes, they differ in their physical formulation and practical implementation.
Synthetic-frequency methods construct the complex-frequency response through frequency-domain synthesis of measured real-frequency data, often incorporating the relevant material-dispersion information.
This strategy provides a rigorous and physically transparent route for emulating complex-frequency excitation and has demonstrated broad applicability across diverse wave systems \cite{guan2023overcoming,zeng2024synthesized,wu2025physics,guan2024compensating,guan2026high}.
TPSF, in turn, offers an alternative temporal-domain formulation based on the Fourier duality between temporal multiplication and spectral convolution.
By compensating effective temporal attenuation in the temporal representation of the measured spectrum, TPSF reconstructs the underlying spectral response without requiring an explicit system-wide dispersion model.
This formulation can be particularly convenient for hybrid plasmonic systems containing multiple materials and resonances.
From a computational perspective, the TPSF workflow provides a simple and experimentally accessible procedure for processing large spectral datasets \cite{khurgin2026revealing}.

\subsection{Broader Implications}
Although the present work focuses on recovering dissipation-obscured Rabi splitting in plasmon--exciton systems, the TPSF framework is more general because it relies on the Fourier relation between temporal attenuation and spectral broadening in linear time-invariant wave systems. The framework may therefore be applicable whenever finite lifetimes and measurement-related broadening conceal physically relevant resonances. Within nanophotonics, potential applications include cavity quantum electrodynamics \cite{imamog1999quantum,chang2014quantum}, exciton--polariton systems \cite{weisbuch1992observation}, photonic-crystal resonators \cite{akahane2003high}, bound states in the continuum \cite{hsu2016bound,reva2026bound}, metasurfaces \cite{yu2014flat}, and non-Hermitian photonics \cite{el2018non}, where finite linewidths can obscure weak resonances, high-$Q$ modes, and subtle modal interactions.
TPSF may consequently benefit spectroscopy \cite{wu2025physics}, sensing \cite{zeng2024synthesized}, and the characterization of dissipative resonant systems \cite{kim2025complex,elbanna20232d,song2024origami,zhang2026far,zhang2025reconstructive,shu2026observation}.
Beyond optical systems, the same Fourier formulation may be extended to microwave, acoustic, elastic, and mechanical resonators.
This broader applicability complements recent developments in complex-frequency excitation \cite{kim2025complex}.
More generally, the space--time Fourier perspective underlying TPSF may provide new insights into temporal wave manipulation in time-varying metasurfaces \cite{yin2022floquet}, where related Fourier relationships have already been experimentally demonstrated \cite{rao2025time,tirole2023double}.


From a fundamental perspective, by introducing TPSF as a formal temporal analogue of the spatial PSF in Fourier optics \cite{goodman1969introduction,braat2008assessment}, this work establishes a unified space--time Fourier framework for spectral reconstruction and offers a different view of dissipation.
Under appropriate conditions, dissipation may be understood not merely as an irreversible degradation of spectral resolution, but as an effective temporal filter that redistributes and encodes recoverable spectral information within the measured response.
The concealed spectral signatures may subsequently be retrieved through inverse temporal reconstruction.
More broadly, this framework provides a general route to accessing spectral features that remain unresolved by conventional measurements.

\section{Conclusions}\label{sec6}

In summary, this work establishes a unified space--time Fourier framework showing that spectral information obscured by dissipation need not be irretrievably lost.
Through the TPSF, effective temporal attenuation and its associated spectral broadening are incorporated into a regularized reconstruction of the underlying spectral response from experimentally measured spectra.
Experiments in deterministic single-molecule AuND--MB nanocavities and open Au@Ag--J-aggregate plasmonic cavities resolve polaritonic branches that are not discernible in the directly measured spectra.
Reconstructions of detuned control systems and compensation-parameter scans further confirm that the recovered mode splitting reflects information already encoded in the measured response rather than an artifact of the reconstruction.
Beyond the plasmon--exciton systems investigated here, this principle may provide a general foundation for loss-aware spectroscopy, sensing, and inverse measurements across photonic, microwave, acoustic, elastic, and mechanical wave systems.


\newpage

\section{Methods}\label{sec11}
\subsection{Optical Simulations}

All optical simulations were performed using the finite-element solver COMSOL Multiphysics 6.5. The experimentally measured dielectric functions of Au and Ag were taken from Johnson and Christy \cite{johnson1972optical}. Perfectly matched layers (PMLs) were applied at the simulation boundaries to eliminate artificial reflections.
For the deterministic single-molecule strong-coupling system, an Au nanosphere dimer was placed on an indium tin oxide (ITO)-coated glass substrate. The hybrid AuND/CB[7]@MB structure was modeled according to the experimentally realized geometry. 
For the Au@Ag nanoparticle systems, both nanorod and nanotriangle geometries were considered. Each nanoparticle consisted of an Au core surrounded by an Ag shell and was placed on a laterally infinite ITO substrate.
The structures were illuminated by a normally incident plane wave, and the scattering spectra, near-field distributions, and scattering cross sections were extracted from the frequency-domain simulations.

\subsection{Fabrication}\label{sec12}

\subsubsection*{Materials}
Gold(III) chloride trihydrate (\ce{HAuCl4.3H2O}, $>99\%$), silver nitrate (\ce{AgNO3}, $99.8\%$), sodium borohydride (\ce{NaBH4}, $99\%$), cetyltrimethylammonium bromide (CTAB, $\geq99\%$), \textit{L}-ascorbic acid (\ce{C6H8O6}, $>99\%$), sodium citrate (\ce{C6H5O7Na3.2H2O}, $99\%$), potassium iodide (\ce{KI}, $99\%$), sodium hydroxide (\ce{NaOH}, $97\%$), cetyltrimethylammonium chloride (CTAC, $99\%$), 1,1$'$-diethyl-2,2$'$-cyanine iodide (PIC, \ce{C23H23IN2}, $99\%$), and sodium chloride (\ce{NaCl}, $99.5\%$) were purchased from Sigma-Aldrich. Sodium 4-(5,6-dichloro-2-(3-(5,6-dichloro-1-ethyl-3-(4-sulfonatobutyl)-1,3-dihydro-2\textit{H}-benzo[d]imidazol-2-ylidene)prop-1-en-1-yl)-1-ethyl-1\textit{H}-benzo[d]imidazol-3-ium-3-yl)butane-1-sulfonate (TDBC, \ce{C29H33Cl4N4NaO6S2}, $98\%$) was purchased from HWRK CHEM. Deionized water with a resistivity of \SI{18.25}{\mega\ohm\centi\meter}, produced using a Millipore Milli-Q purification system, was used throughout.

\subsubsection*{Fabrication of Single-Molecule Plasmon--Exciton Nanocavities}
Single-molecule plasmon--exciton nanocavities were fabricated using the previously reported CB[7]-assisted self-assembly strategy \cite{liu2024deterministic}. Briefly, \SI{5}{\milli\liter} of a \SI{1.0e-7}{\mole} CB[7] aqueous solution was mixed with \SI{2.5}{\milli\liter} of a \SI{1.0e-7}{\mole} methylene blue (MB) solution to form CB[7]@single-MB host--guest complexes. Separately, \SI{4}{\milli\liter} of CTAB-stabilized Au nanoparticle dispersion was centrifuged twice, washed with deionized water, and redispersed in \SI{1}{\milli\liter} of deionized water. Subsequently, \SI{20}{\micro\liter} of the CB[7]@single-MB solution (\SI{5.0e-8}{\mole}) was added to the Au nanoparticle dispersion and incubated for \SI{1}{\hour}.

During self-assembly, the carbonyl-fringed portals of CB[7] bind adjacent Au nanoparticles, enabling the CB[7]@single-MB complexes to function as molecular linkers that bridge individual nanoparticles into Au nanosphere dimers. As a result, a single CB[7]@MB complex is deterministically positioned within the sub-nanometer junction, forming a single-molecule plasmon--exciton nanocavity.

\subsubsection*{Fabrication of Au@Ag Nanocavity J-Aggregate Systems}
The fabrication of Au@Ag nanorods (NRs), Au@Ag nanotriangular prisms (TNPs), and Au@Ag NR/(PIC) J-aggregate structures followed previously reported procedures \cite{liu2017strong,feng2024ultrasensitive}. Here, only the preparation of Au@Ag TNP/(TDBC) J-aggregate nanocavities is briefly described.

TDBC J-aggregates were prepared by dissolving \SI{292}{\milli\gram} of \ce{NaCl} into \SI{1}{\milli\liter} of a \SI{1.25e-5}{\mole} TDBC monomer solution. The mixture was stirred at room temperature in the dark for \SI{20}{\minute}, during which spontaneous self-assembly of TDBC J-aggregates occurred.
To prepare the hybrid Au@Ag TNP/(TDBC) J-aggregate nanocavities, \SI{2.0}{\milli\liter} of the Au@Ag TNP dispersion was centrifuged at \SI{5000}{\per\minute} for \SI{10}{\minute}, washed three times with deionized water, and redispersed in \SI{200}{\micro\liter} of deionized water. Subsequently, \SI{0.8}{\micro\liter} of the TDBC J-aggregate stock solution was added, and the mixture was incubated for \SI{10}{\minute}. Excess J-aggregates and residual TDBC monomers were removed by centrifugation at \SI{4000}{\per\minute} for \SI{8}{\minute}. The resulting Au@Ag TNP/(TDBC) J-aggregate nanocavities were finally redispersed in \SI{3.0}{\milli\liter} of deionized water for subsequent optical characterization.

\subsection{Measurements}

\subsubsection*{Single-Particle Dark-Field Scattering Measurements}
Single-particle dark-field scattering spectra were acquired using an Olympus BX53M microscope (Olympus, Japan) integrated with a monochromator (Acton SpectraPro 2360, Acton Inc., USA), a \SI{150}{\watt} quartz--tungsten--halogen lamp, and a liquid-nitrogen-cooled charge-coupled device (CCD) camera (Princeton Instruments Pixis 100B\_eXcelon, Acton Inc., USA). The CCD detector was operated at \SI{-70}{\degreeCelsius}. Dark-field illumination and backscattered light collection were performed using the same $100\times$ dark-field objective (numerical aperture = 0.90), enabling scattering spectra to be recorded from individual nanocavities. Color dark-field images were obtained using a color digital camera (ARTCAM-300MI-C, ACH Technology Inc.) mounted at the microscope image plane.

For optical measurements, \SI{10}{\micro\liter} of the single-molecule plasmon--exciton nanocavity solution, Au@Ag NR/(PIC) J-aggregate solution, or Au@Ag TNP/(TDBC) J-aggregate solution was drop-cast onto a cleaned substrate and incubated for \SI{3}{\minute}. The substrate was subsequently rinsed with deionized water, dried under a nitrogen flow, and examined by dark-field microscopy. Scattering spectra were then collected from individual nanoparticles.

\subsubsection*{Morphological Characterization}
The morphology of the fabricated nanostructures was characterized using scanning electron microscopy (SEM; Zeiss Auriga-39-34 and JEOL JSM-7610F Plus) operated at an accelerating voltage of \SI{5}{\kilo\volt}. High-resolution structural characterization was performed using transmission electron microscopy (TEM; JEOL JEM-F200) operated at \SI{200}{\kilo\volt}.

\section*{Author contributions}
L.W., C.W.Q., R.L., J.F.Z., and W.Z. conceived the project. J.F.Z. and W.Z. developed the TPSF framework, performed the theoretical analysis and numerical simulations. S.F. and R.L. fabricated the samples and carried out the optical measurements. J.F.Z., W.Z., S.F., Z.S., and R.Z. analyzed and interpreted the data. J.F.Z., W.Z., and L.W. wrote the manuscript with contributions from all authors. R.L. coordinated the experimental work. W.Z., C.W.Q., R.L., and L.W. supervised the project. All authors discussed the results and approved the final manuscript.

\section*{Funding}
This work was supported by the National Research Foundation (NRF), Singapore (Grant Nos. NRF-CRP26-2021-0004 and NRF-CRP31-0007); the Ministry of Education (MOE), Singapore (Grant Nos. MOE-T2EP50223-0001, MOE-MOET32024-0005, and MOE-T2EP50125-0018); the Agency for Science, Technology and Research (A*STAR), Singapore (Grant No. MTC IRG M24N7c0083); and the Singapore University of Technology and Design (SUTD) under the Kickstarter Initiative (Grant No. SKI 2021-04-12). R.L. acknowledges support from the National Natural Science Foundation of China (Grant Nos. 12374326 and 12574388) and the Key Project of the Natural Science Foundation of Henan Province (Grant No. 232300421141).
C.W.Q. acknowledges support from the Ministry of Education, Singapore (Grant Nos. A-8002152-00-00 and A-8002458-00-00); the National Research Foundation, Prime Minister's Office, Singapore, under the Competitive Research Programme (Grant Nos. NRF-CRP26-2021-0004 and NRF-CRP30-2023-0003); and the National Key Research and Development Program of China (Grant No. 2023YFF0613600).

\section*{Data availability}
Data underlying the results presented in this paper are not publicly available at this time but may be obtained from the authors upon reasonable request.

\section*{Conflict of interest}
The authors have no conflict of interest to declare.
\newpage
\bibliography{sn-bibliography}

\newpage
\backmatter



\end{document}